\documentclass[draftclsnofoot,onecolumn]{IEEEtran}
\usepackage{graphicx}
\usepackage{epstopdf} 
\usepackage{verbatim} 
\usepackage{multirow} 
\usepackage{multicol}
\usepackage{amsfonts}
\usepackage{bbm}
\usepackage{subfigure}
\usepackage{colortbl} 
\usepackage{indentfirst}
\usepackage{dcolumn,booktabs}
\usepackage{algorithm}
\usepackage[noend]{algpseudocode}
\usepackage{algorithmicx,algorithm}
\usepackage{setspace}

\usepackage{algpseudocode}
\usepackage{graphics}
\usepackage{epsfig,color}
\usepackage{amsmath}
\usepackage{amssymb}
\ifCLASSINFOpdf

\else
\fi
\newtheorem{Theorem}{Theorem}

\newtheorem{Lemma}{Lemma}

\newtheorem{Remark}{Remark}

\begin{document}
\begin{spacing}{1.0}
\title{Spatio-Temporal Analysis of Cellular Networks with Cell-Center/Edge Users}

\author{
\IEEEauthorblockN{Le Yang, Fu-Chun Zheng and Shi Jin}
}
\maketitle

\begin{abstract}
Emergence of various types of services has brought about explosive growth of traffic as well as diversified traffic characteristics in the cellular networks. To have a comprehensive understanding of the influences caused by various traffic status is vital for the deployment of the next-generation wireless networks. In this paper, we develop a mathematical analytical model by utilizing queuing theory and stochastic geometry where the randomness of the traffic and the geographical locations of the interferers can be captured. We derive the $b$-th moments of the conditional success probability and the closed-form expressions of the meta distribution for the cell-center users (CCUs) and the cell-edge users (CEUs), respectively. Fixed-point equations are then formulated to obtain the exact value of the meta distribution by taking the random arrival traffic into consideration and the impact of the random arrival traffic on the queue status is revealed. In addition, the mean local delays for CCUs nad CEUs are derived and the corresponding regions for CCUs and CEUs where the mean local delays maintain finite are obtained. Finally, the impact of the critical network parameters on the meta distribution and the mean local delay is investigated with the numerical results.
\end{abstract}


%
\IEEEpeerreviewmaketitle

\section{Introduction}
The evolution of mobile applications has led to not only the explosive growth of the data demand \cite{explosive-growth}, but also the the diversified quality of service (QoS) requirements for different types of data traffic due to appearance of new applications, i.e., advanced manufacturing, command-and-control of drones, and tactile Internet \cite{tactile-1}\cite{tactile-2}. The temporal traffic plays a increasingly vital role in the QoS performance of the next generation wireless networks. From the perspective of the network operators, it is essential to attain a comprehensive understanding on the impact of the tremendous traffic and its variations for the deployment of next generation networks \cite{next-generation-networks}.

The broadcast nature of the wireless medium makes the transmitters sharing a common spectrum interact with each other. Moreover, the deployment of the base stations (BSs) introduces the spatial randomness inherently. In order to capture the randomness, stochastic geometry is utilized for the performance evaluation of large-scale networks by modeling the locations of the BSs as a Poisson point process (PPP) \cite{stochastic-geometry-1}-\cite{stochastic-geometry-3}. The analytical tractability and practical relevance into the analysis have popularize its application for performance evaluation among various types of wireless systems, including cellular networks \cite{cellular-networks}, ad-hoc networks \cite{ad-hoc}, heterogeneous networks \cite{heterogeneous-networks}, even with multi-input multi-output (MIMO) technology \cite{MIMO}, and device-to-device (D2D) communication \cite{D2D}. However, in these models, each link is assumed to have full buffer, i.e., there is always a packet to be transmitted for each link. The full buffer assumption restricts these models from being applied to various types of traffic conditions. To overcome the drawback, the additional dimension of randomness is introduced in the temporal domain, i.e., each link between the user and its serving BS is considered as a queue. With this approach, the network performance can also be better understood \cite{traffic-condition-1}\cite{traffic-condition-2}.

As queuing in wireless networks is considered, a natural question is raised that the queues will interact with other and the queue status will be affected by the status of its neighboring queues. The conventional approaches to study the queuing interaction is to utilize the simple collision model \cite{simple-collision-1}-\cite{simple-collision-4}. However, the drawback of this method is obvious that it oversimplifies the wireless channel and hence the effect of interference cannot be tracked. Therefore, queuing theory combined with stochastic geometry is utilized to characterize the randomness for both temporal and spatial domains \cite{combine-1}-\cite{combine-3}. In \cite{combine-4}, the sufficient and necessary conditions for network stability were provided. Different performance metrics, such as successful transmission probability \cite{combine-1} and throughput \cite{combine-2}\cite{combine-3}, were subsequently derived. However, the results in \cite{combine-2} are only applicable to light traffic condition and the bounds given in \cite{combine-4} are not necessary tight. In addition, \cite{combine-1} accounted the success transmission by using the spatially averaged performance and an accurate approximation cannot always be provided.

To provide a detailed characterization on the coupling between the network service and the data traffic, Meta distribution is introduced in \cite{meta-distribution}. The success probability corresponds to the link reliability, which answers the question: ``What fraction of users can receive successful transmission?'' while the meta distribution provides the fine-grained latency and reliability analysis for wireless networks and gives answer to the question ``What fraction of users can achieve $x$\% reliability?'' Clearly, the success probability only characterize the mean of the success probability for all users while a considerable amount of valuable information cannot be provided. We take the comparison of two scenarios for example. One scenario is that 50$\%$ of the users attain 8$\%$ reliability and 50$\%$ of the users attain 94$\%$ reliability. The other is that all users attain 51$\%$ reliability. The success probabilities for two scenarios are both 51$\%$. We can see that there are huge differences in two scenarios in terms of QoS which cannot be reflected in the conventional performance metric while this difference can be revealed by the meta distribution. Moreover, other critical network performance measures, i.e., the mean local delay, the network jitter and the variance of the success probability can also be captured by utilizing the meta distribution. Recent works focus on the performance evaluation by utilizing the meta distribution in various wireless systems, including bipolar networks \cite{meta-distribution}, cellular networks \cite{md-cellular-networks}, heterogeneous networks \cite{md-hetnet}, even with cell-range expansion, D2D communication \cite{md-D2D}, BS cooperation \cite{md-bs-cooperation}, non-orthogonal multiple access (NOMA) \cite{md-NOMA} and power control \cite{md-power-control}. In \cite{JSAC}, the delay in the heterogeneous networks with random arrival of traffic was analyzed and the statistics of the spatio-temporal traffic was evaluated. In addition, the lower and upper bounds of delay distribution is derived in different scheduling schemes. However, the analysis does not capture the effect of queuing interaction on the wireless networks and the bound can be loose.

Note that the queue status for a link is affected by its surrounding environments, i.e., the distance of the link as well as the queue status and the distribution of its neighbors. That is, the link quality is heavily dependent on its location as the cell-edge users (CEU) receive weaker signal from its serving BS and stronger interference from the other BSs, leading to the degradation of the success probability and achievable rate. In contrast, for the cell-center users (CCU), the interference is significantly lower than the received signal. Therefore, CCU can experience larger success probability than CEU. The classification of user is studied in \cite{cr-grid} for the grid-based networks. For the PPP-model networks, the instantaneous SINR-based criterion is utilized to categorize CCU and CEU in \cite{cr-SINR-1} and \cite{cr-SINR-2} and the distance-based criterion is adopted in \cite{cr-distance}\cite{load-aware}.

Motivated by the above, we provide a more fine-grained analysis for cellular networks by categorizing the users into CCUs and CEUs based on their location. In this paper, the spatio-temporal traffic is modeled by utilizing the tools from queuing theory and stochastic geometry. Specifically, the spatial distribution of users is modeled by PPP and the arrival of packets at each user is modeled by independent Bernoulli processes. With the developed framework, the temporal variation of the traffic and its impact on system stability can be characterized. Our main contribution can be summarized as follows:
\begin{enumerate}
\item We present a tractable analytical framework that captures the coupling among the spatial distribution of wireless links and the temporal traffic dynamic. In addition, the users are classified into CEUs and CCUs based on the locations within a cell. The critical features of a cellular network, i.e., small-scale fading and path loss, traffic profile and spatial randomness of BSs are also taken into consideration. With the framework, the spatial and temporal characteristics for cellular network performance can be analyzed based on the user classification.
\item We derive the $b$-th moments of the conditional success probability for CCU and CEU by taking the active probability into consideration, respectively. The expression of meta distributions for CCU and CEU and their Beta approximations are obtained by utilizing the derived results. In addition, the mean local delays for CCU and CEU are derived. A fixed-point equation is proposed to identify the critical value of the mean local delay under each user classification.
\item  The fraction of CCUs and CEUs achieving an arbitrary success probability is the function of the critical spatial and temporal parameters. A fixed-point equation is formulated to obtain the exact value of the fraction of users achieving arbitrary level of success probability for a given arrival rate. An iterative method is provided in the static regime for the cellular networks and a simplified fixed-point equation is proposed to obtain the results in the special case where the cellular network is in the high mobility regime. In addition, the necessary and sufficient conditions for the network stability are given.
\item By numerical simulations, the effects of critical network parameters on the meta distribution is revealed. In addition, the mean local delay increases with the active probability. A CCU has a larger local delay than a CEU while the critical value for a CCU is much larger than the a CEU.
\end{enumerate}

The reminder of this paper is organized as follows. In Section $\text{\uppercase\expandafter{\romannumeral2}}$, the system model is introduced. In Section $\text{\uppercase\expandafter{\romannumeral3}}$, some auxiliary results are provided. In Section $\text{\uppercase\expandafter{\romannumeral4}}$, the meta distribution analysis is provided. In Section $\text{\uppercase\expandafter{\romannumeral5}}$, numerical results are presented to demonstrate the impact of network parameters. Finally, conclusions are provided in Section $\text{\uppercase\expandafter{\romannumeral6}}$.

\section{System model}
\begin{figure}
  \centering
  \includegraphics[width=3.5in]{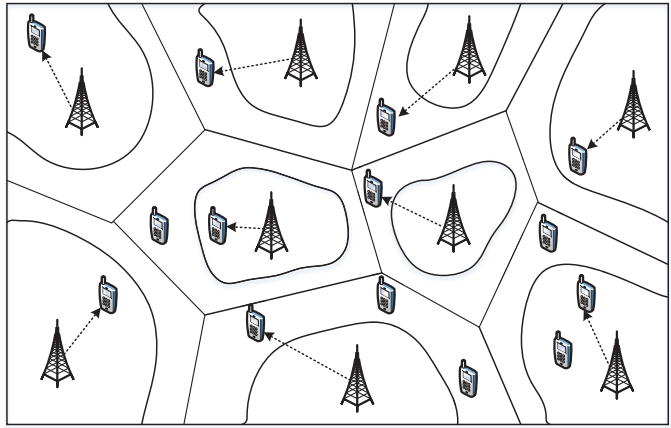}
  \caption{A cellular network consisting of CCUs and CEUs. The users located within the irregular circle of Voronoi cells are CCUs. The users located outside the irregular circle are CEUs.}\label{framework-CCU-CEU}
\end{figure}

We consider a downlink cellular network. The BSs and the users are assumed to be distributed as a PPP with density $\lambda$ and $\lambda_u$, respectively. The users are classified into two sets, i.e., the CCUs and the CEUs, as shown in Fig. \ref{framework-CCU-CEU}. The channel gain between the users and the BSs is modeled by the small-scale fading and the large-scale fading. The small-scale fading coefficient is assumed to be an exponential random variable with unit mean (Rayleigh fading). A path loss function $\ell(r)=r^{-\alpha}$ is adopted, where $\alpha$ is the path loss exponent and $r$ the distance between the serving BS and the user. The BSs transmit with the constant power $P$. In order to study the temporal variation of the traffic, the mobility of the users are limited and the network is considered to be static. Without loss of generality, according to Slivnyak's theorem \cite{slivnyak}, we study the performance of a typical user $u_0$ located at the origin $o\in\mathbb{R}$.

The time is assumed to be segmented into time slots with equal durations. All events in this discrete time queuing system, i.e., the arrival and departure of the packets, occur around the boundaries of time slots. In other words, the arrivals of the packets occur at the instant immediately after the boundaries while the departure of the packets occur at the instant immediately before the boundaries. For each user located within the Voronoi cell of its tagged BS in $\Phi$, the arrival process of the packets is modeled as a independent Bernoulli process with rate $\xi\in[0,1]$, indicating that the packets arrive at the user with probability $\xi$ in any time slot. We assume that the size of all packets are equal and one time slot is required to transmit a packet. Moreover, we assume each BS maintains an infinite-size queue to store the incoming packet for each user located within its coverage.

We adopted the association criterion where each user is associated with its closest BS. Under this criterion, multiple users can be served by a BS. In order to avoid contention between different users, the random scheduling scheme is adopted. Specifically, since each queue in the BS corresponds to an individual user in the coverage of the BS, the BS randomly chooses a non-empty queue in each time slot and transmits the head-of-line packet to the corresponding user. If all the queues in the BS are empty at a time slot, the BS will mute its transmission to reduce the inter-cell interference as well as the power consumption.

\begin{figure}
  \centering
  \includegraphics[width=3.5in]{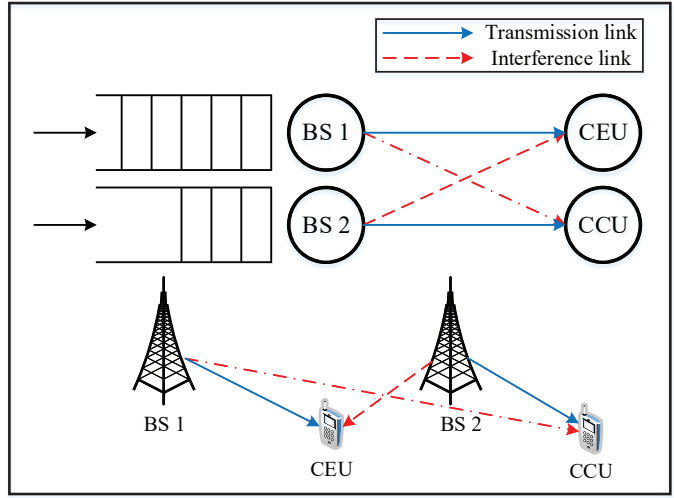}
  \caption{Illustration of queue interaction.}\label{interacting-queues}
\end{figure}

We consider a interference-limited cellular network where the thermal noise is ignored. A transmission process is considered to be successful when the received SIR at the user exceeds a predefined threshold $\theta$ and the feedback of the transmission, i.e., success or failure, can be aware by the BS immediately. If the transmission succeeds, the packet will be removed out of the queue. Otherwise, the packet will be added to the head of the corresponding queue and wait to be retransmitted. From the transmission process, it can be observed that whether the BS is transmitting or not is dependent on the queue status and the scheduling scheme at time slot $t$. Let $\zeta_{x,t}\in\{0,1\}$ be an indicator showing that whether the BS at $x\in\Phi$ is transmitting ($\zeta_{x,t}=1$) or not ($\zeta_{x,t}=0$) at time slot $t$. When $u_0$ is associated with a BS, the received signal-to-interference-ratio (SIR) at $u_0$ can be written as follows
\begin{equation}\label{SIR}
\text{SIR}_c=\frac{P|h_{x_0}|^2x_0^{-\alpha}}{I_{c}},\ \ \text{and}\ \ \text{SIR}_e=\frac{P|h_{x_0}|^2x_0^{-\alpha}}{I_{e}},
\end{equation}
where $I_c=I_e=\sum_{x\in\Phi\backslash x_0}\zeta_{x,t}P|h_{x}|^2x^{-\alpha}$. From (\ref{SIR}), it can be observed that the SIR affect, and be affected, by the queue statuses at all BSs. Note that the indicator $\zeta_{x,t}$ is both temporal and spatial dependent since the queue size at each BS varies with time and the location affects the aggregated interference. We provided a simple scenario in Fig. \ref{interacting-queues} to illustrate the queue interaction between two BSs. BS 1 is serving a CEU and the distance of the transmission link is comparable to that of the interference link. Meanwhile, BS 2 is serving a CCU and the distance of the transmission link is much smaller than that of the interference link. Therefore, the CCU receives fewer interference and smaller path loss, resulting in that the CCU can empty its queue quickly. The disparity between their transmission condition leads to that the CEU actives more frequently than the CCU. Moreover, whether the packets arrive at both BSs or not will make a difference on the active duration for both BSs: If the packets appear only in one BS, the other BS will benefit from the smaller interference and the queue flushing process can be sped up. Therefore, the active duration is decreased. Otherwise, the service rate decreases with the mutual interference and the active duration is prolonged. To this end, the coupling between the SIR and the queue statuses at all BSs makes the analysis difficult. In order to shed light on the effect of the queue status on the network performance, we first assume that $\zeta_{x,t}$, which is both temporal and spatial independent, is a constant, i.e.,
\begin{equation}
\mathcal{P}\left(\zeta_{x,t}=1\right)=q,
\end{equation}
and derive the closed-form expression for the meta distribution, then obtain the exact form of the meta distribution by taking into account the randomness of the arrival traffic.

The meta distribution $\bar{F}_{\mathcal{P}}(x)$ is defined as the complementary cumulative distribution function (CCDF) of the conditional success probability
\begin{equation}
\mathcal{P}(\theta)=\mathbbm{P}(\text{SIR}>\theta|\Phi),
\end{equation}
which is the CCDF of the SIR for $u_0$ conditioned on the realization of $\Phi$. Henceforth, the meta distribution is given by \cite{meta-distribution}
\begin{equation}
\bar{F}_{\mathcal{P}}(x)\triangleq\mathbbm{P}(\mathcal{P}(\theta)>x),\ x\in[0,1].
\end{equation}

Due to the ergodicity of the point processes, the meta distribution can be regarded as the fraction of active links with the conditional success probability greater than $x$.

Since the direct calculation of the meta distribution is infeasible, it is essential for the derivation of meta distribution to obtain the $b$-th moment of the conditional success probability. Denoting the $b$-th moment of $\mathcal{P}$ by $M_b$, we can derive the expression of meta distribution by utilizing the Gil-Pelaez theorem. In addition, a simpler alternative method is adopted to utilize the beta distribution to approximate the meta distribution by matching the first and second moments.

We also analyze the per-link delay consisting of two parts, i.e., the queuing delay and the transmission delay. The transmission delay mainly consists of waiting delay and retransmission delay. This type of delay can also be called the local delay, which is defined as the number of retransmissions needed until a successful transmission occurs \cite{md-D2D}. We denoted the local delay by $L$ and thus the mean local delay is written as
\begin{equation}\label{delay-definition}
\mathbbm{E}[L]\overset{(a)}{=}\mathbbm{E}\left[\frac{1}{\mathcal{P}(\theta)}\right]=M_{-1}.
\end{equation}
where (a) follows from the fact that $L$ is geometrically distributed with parameter $\mathcal{P}(\theta)$ conditioned on $\Phi$. Denoting $L|\Phi$ by $L_{\Phi}$, we have
\begin{equation}
\mathbbm{P}(L_{\Phi}=k)=(1-\mathcal{P})^{k-1}\mathcal{P},\ k\in\mathbbm{N}.
\end{equation}
where $\mathcal{P}$ is the conditional success probability. As shown in (\ref{delay-definition}), the mean local delay can be derived by computing the $-1$-st moment of the conditional success probability.

\section{Auxiliary Results}
In this section, we first derive the probability for $u_0$ to be CCU or CEU. $u_0$ is termed as a CCU if $\frac{R_m}{R_d}<R$ and otherwise as a CEU, where $R_m$ denotes the distance between $u_0$ and its serving BS, $R_d$ the distance between $u_0$ and its dominant interfering BS (second nearest BS) and $R$ a predefined ratio threshold. According to \cite{distance-distribution}, the joint distribution of $R_m$ and $R_d$ is give by
\begin{equation}\label{joint-distribution}
f_{R_m,R_d}(r_m,r_d)=(2\pi\lambda)^2r_mr_d\exp\left(-\pi\lambda r_d^2\right)
\end{equation}

By using (\ref{joint-distribution}), the probability of $u_0$ being CCU can be derive as follows
\begin{equation}\label{ccu-probability}
\mathbbm{P}\left[\frac{R_m}{R_d}<R\right]=\int_{0}^{\infty}\int_{0}^{r_dR}f_{R_m,R_d}(r_m,r_d)\text{d}r_d\text{d}r_m=R^2
\end{equation}

Accordingly, the probability of $u_0$ being CEU is $1-R^2$.

Next, we provide some auxiliary results which are essential for the derivation of the meta distribution, including the CCDF and the probability density function (PDF) of the distance between $u_0$ and its serving BS.

Let $R_c$ and $R_d$ be the distance of a CCU from its serving BS and the dominant interfering BS. Therefore, the CCDF of the distance between CCU and its serving is given by
\begin{equation}
\bar{F}_{R_c}=\exp\left(-\pi\lambda\frac{r_c^2}{R^2}\right)
\end{equation}

\emph{Proof:}
Conditioning on $u_0$ being a CCU, the probability that $R_c$ is larger than $r_c$ can be derived as follows
\begin{equation}
\begin{split}
\bar{F}_{R_c}&=\mathbbm{P}\left[R_c>r_c|R_c\leq R_dR\right]\\
&=\frac{\mathbbm{P}\left[R_c>r_c,R_c\leq R_dR\right]}{\mathbbm{P}\left[R_c\leq R_dR\right]}\\
&=\frac{1}{R^2}\int_{r_c}^{\infty}\int_{r_c/R}^{\infty}f_{R_c,R_d}(r_c,r_d)\text{d}r_d\text{d}r_c\\
&=\exp\left(-\pi\lambda\frac{r_c^2}{R^2}\right)
\end{split}
\end{equation}

The PDF of the distance between a CCU and its serving BS can be derived as follows
\begin{equation}\label{ccu-pdf}
f_{R_c}(r_c)=-\frac{\text{d}\bar{F}_{R_c}(r_c)}{\text{d}r_c}=2\pi\lambda\frac{r_c}{R^2}\exp\left(-\pi\lambda\frac{r_c^2}{R^2}\right)
\end{equation}

Let $R_e$ be the distance between a CEU and its serving BS. The CCDF of $R_e$ is given by
\begin{equation}
\bar{F}_{R_e}=\frac{1}{1-R^2}\left(\exp\left(-\pi\lambda r_e^2\right)R^2-\exp\left(-\pi\lambda\frac{r_e^2}{R^2}\right)\right)
\end{equation}

\emph{Proof:}
Conditioning on $u_0$ being a CEU, the probability that $R_e$ is larger than $r_e$ can be derived as follows
\begin{equation}
\begin{split}
\bar{F}_{R_e}&=\mathbbm{P}\left[R_e>r_e|R_e\leq R_dR\right]\\
&=\frac{\mathbbm{P}\left[R_e>r_e,R_e\leq R_dR\right]}{\mathbbm{P}\left[R_e\leq R_dR\right]}\\
&=\frac{1}{R^2}\int_{r_e}^{\infty}\int_{r_e/R}^{\infty}f_{R_e,R_d}(r_e,r_d)\text{d}r_d\text{d}r_e\\
&=\frac{1}{1-R^2}\left(\exp\left(-\pi\lambda r_e^2\right)R^2-\exp\left(-\pi\lambda\frac{r_e^2}{R^2}\right)\right)
\end{split}
\end{equation}

Accordingly, the PDF of $R_e$ can be derived as follows
\begin{equation}\label{ceu-pdf}
\begin{split}
f_{R_e}(r_e)&=-\frac{\text{d}\bar{F}_{R_e}(r_e)}{\text{d}r_e}\\
&=\frac{2\pi\lambda r_e}{1-R^2}
\left(\exp\left(-\pi\lambda r_e^2\right)-\exp\left(-\pi\lambda\frac{r_e^2}{R^2}\right)\right)
\end{split}
\end{equation}
\section{Meta Distribution}
In this section, we derive the $b$-th moments of the conditional success probability for CCU and CEU, i.e., $M_{b,c}$ and $M_{b,e}$, followed by the exact expression and closed-form approximations of the meta distribution for CCU and CEU, respectively. In order to show  the impact of the active probability $q$ on the meta distribution, we derive the meta distribution of the conditional success probability conditioned on that each interfering BS is active with probability $q$ independently, then provide a fixed-point equation to obtain the exact expression of the meta distribution when the active probability is considered to be a temporal and spatial dependent random variable.

\subsection{Meta Distribution for CCU}
\begin{figure}
  \centering
  \includegraphics[width=3.5in]{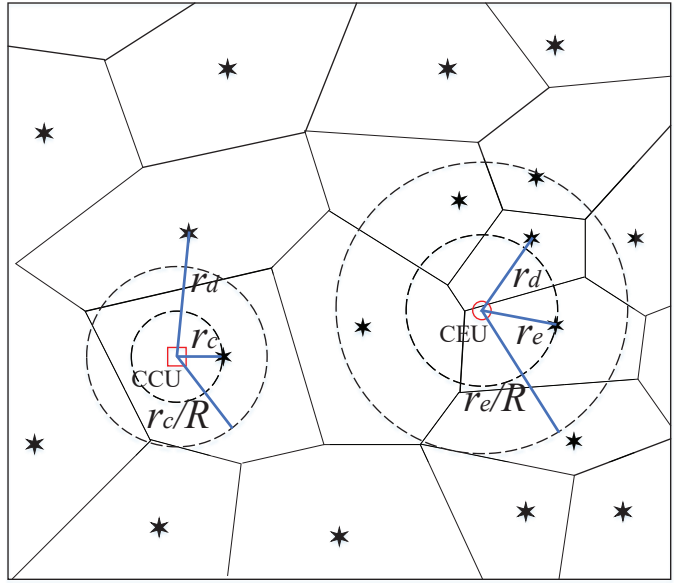}
  \caption{Voronoi tessellation for CCU and CEU.}\label{Voronoi-tessellation}
\end{figure}

From Fig. \ref{Voronoi-tessellation}, it can be observed the serving BS is at $r_c$ and the dominant interfering BS is at $r_d$. The dominant interfering BS lies beyond the circle with radius $\frac{r_c}{R}$. As each BS is active independently, the locations of the active BSs are modeled as a thinned PPP with density $q\lambda$. We first provide the $b$-the moment of $\mathcal{P}_c$ in the following theorem, for any $b\in\mathbbm{C}$.
\begin{Theorem}
The $b$-the moment of conditional success probability for CCU is given by
\begin{equation}\label{ccu-moment}
\begin{split}
M_{b,c}&=\left(1+\delta\sum_{n=1}^{\infty}\binom{b}{n}(-1)^{n+1}\frac{(q\theta)^{n}R^{\alpha n}}{n-\delta}\right.\\
&\left.{}_2F_1\left(n,n-\delta;n-\delta+1;-\theta R^{\alpha}\right)\right)^{-1}
\end{split}
\end{equation}
\end{Theorem}
\emph{Proof:} See the Appendix A.

By applying the Gil-Pelaez theorem, the meta distribution of the SIR for a CCU is given by
\begin{equation}\label{ccu-md}
\bar{F}_{\mathcal{P}_c}=\frac{1}{2}+\frac{1}{\pi}\int_{0}^{\infty}\frac{\mathcal{J}\left(e^{-jt\log x}M_{jt,c}\right)}{t}\text{d}t,
\end{equation}
where $\mathcal{J}(z)$ is the imaginary part of $z\in\mathbbm{c}$. Since the numerical evaluation of (\ref{ccu-md}) is cumbersome and it is difficult to obtain further insight, we resort to a beta distribution to approximate the meta distribution by matching the first and second moments, which can be easily obtained from the result in (\ref{ccu-moment}):
\begin{equation}
M_{1,c}=\left(1+\frac{q\theta\delta R^{\alpha}}{1-\delta}{}_{2}F_1\left(1,1-\delta;2-\delta;-\theta R^{\alpha}\right)\right)^{-1}
\end{equation}
\begin{equation}
\begin{split}
M_{2,c}=&\left(1+\frac{2q\theta\delta R^{\alpha}}{1-\delta}{}_{2}F_1\left(1,1-\delta;2-\delta;-\theta R^{\alpha}\right)\right.\\
&\left.-\frac{(q\theta)^2\delta R^{2\alpha}}{2-\delta}{}_{2}F_1\left(2,2-\delta;3-\delta;-\theta R^{\alpha}\right)\right)^{-1}
\end{split}
\end{equation}

By matching the variance and mean of the beta distribution, i.e., $M_{2,c}-M_{1,c}^2$ and $M_{1,c}$, the approximated meta distribution of the SIR for a CCU can be given by
\begin{equation}\label{ccu-md-approximation}
\bar{F}_{\mathcal{P}_c}\approx 1-I_x\left(\frac{M_{1,c}}{1-M_{1,c}},\beta\right),\ x\in[0,1],
\end{equation}
where
\begin{equation}
\beta=\frac{(M_{1,c}-M_{2,c})(1-M_{1,c})}{M_{2,c}-M_{1,c}^2}
\end{equation}
and $I_x(a,b)$ is the regularized incomplete beta function
\begin{equation}
I_x(a,b)\triangleq\frac{\int_{0}^{x}t^{a-1}(1-t)^{b-1}\text{d}t}{\text{B}(a,b)}.
\end{equation}

From (\ref{ccu-md-approximation}) and (\ref{ccu-md}), we can see that the meta distribution depends on the active probability $q$. In practice, $q$ is jointly affected by the arrival rate $\xi$ and the success probability. Since the distributions of the success probability for CCUs and CEUs are different, the coupling between the CCUs and CEUs makes the derivation of the active probability difficult. Therefore, we resort to the scenario where all uses in the cellular networks are in the cell-center region. We denote by $\nu$ the number of users served by each BS. By utilizing the Little's law \cite{little}, the active probability of a CCU is given by
\begin{equation}\label{active-probability}
q=
\begin{cases}
1, &\text{if}\ \mathcal{P}_c\leq\nu\xi\\
\frac{\nu\xi}{\mathcal{P}_c}, &\text{if}\ \mathcal{P}_c>\nu\xi
\end{cases}
\end{equation}

Next, we derive the mean active probability of the cellular network. According to \cite{PMF}, the probability mass function (PMF) of the number of the users is given by
\begin{equation}
g_{\text{N}}(\nu)=\frac{3.5^3.5\Gamma(\nu+3.5)\left(\frac{\lambda_u}{\lambda}\right)^{\nu}}{\nu!\Gamma(3.5)
\left(\frac{\lambda_u}{\lambda}+3.5\right)^{\nu+3.5}}
\end{equation}

Note that only the BSs with users in its Voronoi cell is considered since they have packets to send. Moreover, the distribution of the success probability can be characterized by the meta distribution, we then have the mean active probability for a CCU as follows
\begin{equation}
\mathbbm{E}(q)=\sum_{\nu=1}^{\infty}\frac{g_{\text{N}}(\nu)}{1-g_{\text{N}}(0)}
\left(1-\bar{F}_{\mathcal{P},c}(\nu\xi)-\int_{\nu\xi}^{1}\frac{\nu\xi}{s}\bar{F}_{\mathcal{P},c}(\text{d}s)\right)
\end{equation}

Based on the above analysis, a fixed-point equation can be formulated to obtain the exact meta distribution in the following lemma.
\begin{Lemma}
The meta distribution can be obtained by solving the fixed-point equation (\ref{fixed-point}), shown at the top of the next page, where $\text{Im}\{\cdot\}$ denotes the imaginary part of a complex number.
\begin{figure*}
\begin{equation}\label{fixed-point}
\begin{split}
\bar{F}_{\mathcal{P},c}=&\frac{1}{2}+\frac{1}{\pi}\int_{0}^{\infty}\frac{1}{t}\text{Im}\left(e^{-jt\log x}
\left(1+\delta\sum_{n=1}^{\infty}\binom{b}{n}(-1)^{n+1}\sum_{\nu=1}^{\infty}\frac{g_{\text{N}}(\nu)}{1-g_{\text{N}}(0)}
\left(1-\bar{F}_{\mathcal{P},c}(\nu\xi)-\int_{\nu\xi}^{1}\frac{(\nu\xi)^n}{s^n}\bar{F}_{\mathcal{P},c}(\text{d}s)\right)\right.\right.\\
&\left.\left.\frac{\theta^{n}R^{\alpha n}}{n-\delta}{}_2F_1\left(n,n-\delta;n-\delta+1;-\theta R^{\alpha}\right)\right)^{-1}
\right)\text{d}t
\end{split}
\end{equation}
\end{figure*}
\end{Lemma}

\begin{Remark}
When all the users are located within cell-center region, the users tend to be influenced by each other due to their strong ability to diminishing the queue. Hence, the derived results provide a lower bound of the network performance. In the scenario where different frequency bands are reserved for CCUs and CEUs. According to the derived results in (\ref{ccu-probability}), the CEUs are distributed as a thinned PPP with density $qR^2$ \cite{stochastic-geometry-1}. By replacing $q$ with $qR^2$ in (\ref{ceu-md}), we can obtain the meta distribution for the CEUs in the scenario where different frequency bands are reserved for CCUs and CEUs. In addition, it is difficult to obtain the results due to the complicated form of the fixed-point equation. Therefore, an recursive approach is adopted to obtain a tight approximation of the meta distribution. Note that this approach depends on the assumption of the independence of the transmission success events between different time slots. The independence can be guaranteed since the temporal correlation is reduced significantly due to the independent fading and the employment of random scheduling scheme. The approach is given by
\begin{equation}
\bar{F}_{\mathcal{P},c}=\lim_{t\rightarrow\infty}\bar{F}_{\mathcal{P},c,t}.
\end{equation}

The basic idea of this approach is to utilize the meta distribution of time slot $t-1$ to obtain the active probability of time slot $t$. The initial value of active probability is equal to the arrival rate $\zeta$ as all the queues are empty at time slot 0. A tight approximation can be obtained when $t\rightarrow\infty$.
\end{Remark}

\subsection{Meta Distribution for CEU}
As the distance of a CEU from the serving BS and the dominant interfering BS is relatively close, whether the dominant interfering BS is in the active mode has a significant impact on the QoS experienced by a CEU. From Fig. \ref{Voronoi-tessellation}, it can be observed the serving BS is at $r_e$ and the dominant interfering BS is at $r_d$. The dominant interfering BS lies in the ring formed by two circles with radius $r_e$ and $\frac{r_e}{R}$. As each BS is active independently, the locations of the active BSs are modeled as a thinned PPP with density $q\lambda$. In addition, the dominant interfering BS persists in the thinned process of the active BSs with probability $q$ and do not persist in the thinned process with probability $1-q$. We then derive the $b$-th moment of the conditional success probability in two cases.

\emph{Case 1 (dominant interfering BS is active):} In this case, the dominant interfering BS is in the active mode. That is, at least one interfering BS exists in the ring formed by two circles with radius $r_e$ and $r_e/R$. Therefore, the active BSs are split into two sets $\mathcal{S}_1(r_e)$ and $\mathcal{S}_2(r_e)$. The set $\mathcal{S}_1(r_e)$ is defined as $\mathcal{S}_1(r_e)=\{x\in\Phi(t)|r_e\leq\|x\|\leq \frac{r_e}{R}\}$ and the set $\mathcal{S}_2(r_e)$ is defined as $\mathcal{S}_2(r_e)=\{x\in\Phi(t)\||x\|\geq \frac{r_e}{R}\}$. Therefore, the set of active BSs for a CEU follows a PPP with zero density in $B(0,r_e)$ and density $q\lambda$ in $B(0,\frac{r_e}{R})\backslash B(0,r_e)$ (conditioned on the existence of at least one point) and $\mathbbm{R}^2\backslash B(0,\frac{r_e}{R})$. Accordingly, the $b$-the moment of the conditional success probability can be provided in the following lemma.
\begin{Lemma}
Conditioned on that the dominant interfering BS is active, the $b$-the moment of the conditional success probability for a CEU is given by
\begin{equation}\label{M-e1}
\begin{split}
M_{b,e1}=&\frac{R^2}{1-R^2}\sum_{t=0}^{\infty}\sum_{l=0}^{1}\sum_{m=0}^{1}(-1)^{l+m+1}\left((t+m)(1-R^2)\right.\\
&\left.+|m-1|R^2V_1(\theta)+(1-|m-1|)V_2(\theta)+R^{2l}\right)^{-1},
\end{split}
\end{equation}
where
\begin{equation}\label{V-1}
\begin{split}
&V_1(\theta,b)=\\
&\delta\sum_{n=0}^{\infty}\binom{b}{n}(-1)^{n+1}\frac{(q\theta)^n}{n-\delta}{}_{2}F_1\left(n,n-\delta;n-\delta+1;-\theta\right)
\end{split}
\end{equation}
\begin{equation}\label{V-2}
\begin{split}
&V_2(\theta,b)=\\
&\delta\sum_{n=0}^{\infty}\binom{b}{n}(-1)^{n+1}\frac{(q\theta)^nR^{\alpha n}}{n-\delta}{}_{2}F_1\left(n,n-\delta;n-\delta+1;-\theta R^{\alpha}\right)
\end{split}
\end{equation}
\end{Lemma}
\emph{Proof:} See the Appendix B.

\emph{Case 2 (dominant interfering BS is not active):} In this case, the condition that at least one interfering BS exists in the set $\mathcal{S}(r_e)$ is relaxed. Hence, we straightforwardly focus on the set of interfering BSs which includes all BSs beyond the distance $r_e$. In other words, the set of active BSs for a CEU follows a PPP with zero density in $B(0,r_e)$ and density $q\lambda$ in $\mathbbm{R}^2\backslash B(0,r_e)$.

\begin{Lemma}
Conditioned on that the dominant interfering BS is not active, the $b$-the moment of the conditional success probability for a CEU is given by
\begin{equation}\label{M-e2}
\begin{split}
M_{b,e2}=&\frac{1}{1-R^2}\left(\frac{1}{1+V_1(\theta)}-\frac{R^2}{1+R^2V_1(\theta)}\right),
\end{split}
\end{equation}
where $V_1(\theta)$ and $V_2(\theta)$ are given by (\ref{V-1}) and (\ref{V-2}), respectively.
\end{Lemma}
\emph{Proof:} See the Appendix C.

Combining (\ref{M-e1}) and (\ref{M-e2}), the $b$-th moment of the conditional success probability for a CEU can be expressed as follows
\begin{equation}
\begin{split}
&M_{b,e}=qM_{b,e1}+(1-q)M_{b,e2}\\
=&\frac{qR^2}{1-R^2}\sum_{t=0}^{\infty}\sum_{l=0}^{1}\sum_{m=0}^{1}(-1)^{l+m+1}\left((t+m)(1-R^2)\right.\\
&\left.+|m-1|R^2V_1(\theta,b)+(1-|m-1|)V_2(\theta,b)+R^{2l}\right)^{-1}\\
&+\frac{1-q}{1-R^2}\left(\frac{1}{1+V_1(\theta,b)}-\frac{R^2}{1+R^2V_1(\theta,b)}\right)
\end{split}
\end{equation}

By applying the Gil-Pelaez theorem, the meta distribution of the SIR for a CCU is given by
\begin{equation}\label{ceu-md}
\bar{F}_{\mathcal{P}_e}=\frac{1}{2}+\frac{1}{\pi}\int_{0}^{\infty}\frac{\mathcal{J}\left(e^{-jt\log x}M_{jt,e}\right)}{t}\text{d}t,
\end{equation}

The first moment of the conditional success probability can be expressed as follows
\begin{equation}\label{M-e-1}
\begin{split}
M_{1,e}&=\frac{qR^2}{1-R^2}\sum_{t=0}^{\infty}\sum_{l=0}^{1}\sum_{m=0}^{1}(-1)^{l+m+1}\left((t+m)(1-R^2)\right.\\
&\left.+|m-1|R^2V_1(\theta,1)+(1-|m-1|)V_2(\theta,1)+R^{2l}\right)^{-1}\\
&+\frac{1-q}{1-R^2}\left(\frac{1}{1+V_1(\theta,1)}-\frac{R^2}{1+R^2V_1(\theta,1)}\right)
\end{split}
\end{equation}

Similarly, the second moment of the conditional success probability is written as follows
\begin{equation}\label{M-e-2}
\begin{split}
M_{2,e}&=\frac{qR^2}{1-R^2}\sum_{t=0}^{\infty}\sum_{l=0}^{1}\sum_{m=0}^{1}(-1)^{l+m+1}\left((t+m)(1-R^2)\right.\\
&\left.+|m-1|R^2V_1(\theta,2)+(1-|m-1|)V_2(\theta,2)+R^{2l}\right)^{-1}\\
&+\frac{1-q}{1-R^2}\left(\frac{1}{1+V_1(\theta,2)}-\frac{R^2}{1+R^2V_1(\theta,2)}\right),
\end{split}
\end{equation}

By utilizing (\ref{M-e-1}) and (\ref{M-e-2}), the approximated meta distribution of the SIR for a CEU can be given by
\begin{equation}
\bar{F}_{\mathcal{P}_e}\approx 1-I_x\left(\frac{M_{1,e}}{1-M_{1,e}},\beta\right),\ x\in[0,1],
\end{equation}
where
\begin{equation}
\beta=\frac{(M_{1,e}-M_{2,e})(1-M_{1,e})}{M_{2,e}-M_{1,e}^2}
\end{equation}

Similar with the derivation for CCUs, the coupling between the CCUs and CEUs makes the derivation of the active probability difficult. Therefore, we resort to the scenario where all uses in the cellular networks are in the cell-edge region. We denote by $\nu$ the number of users served by each BS. By utilizing the Little's law \cite{little}, the active probability of a CCU is given by
\begin{equation}\label{active-probability}
q=
\begin{cases}
1, &\text{if}\ \mathcal{P}_e\leq\nu\xi\\
\frac{\nu\xi}{\mathcal{P}_e}, &\text{if}\ \mathcal{P}_c>\nu\xi
\end{cases}
\end{equation}

The mean active probability for a CEU is given by
\begin{equation}\label{ceu-mean-active}
\mathbbm{E}(q)=\sum_{\nu=1}^{\infty}\frac{g_{\text{N}}(\nu)}{1-g_{\text{N}}(0)}
\left(1-\bar{F}_{\mathcal{P},e}(\nu\xi)-\int_{\nu\xi}^{1}\frac{\nu\xi}{s}\bar{F}_{\mathcal{P},e}(\text{d}s)\right)
\end{equation}

Based on the above analysis, a fixed-point equation can be formulated to obtain the exact meta distribution for a CEU in the following lemma.
\begin{Lemma}
The fixed-point equation is obtained by substituting (\ref{ceu-mean-active}) into (\ref{fixed-point}), where $V_1(\theta,b)$ and $V_2(\theta,b)$ are given by (\ref{v1}) and (\ref{v2}), as shown at the top of the next page.
\begin{figure*}
\begin{equation}\label{v1}
\begin{split}
V_1(\theta,b)=\delta\sum_{n=0}^{\infty}\binom{b}{n}(-1)^{n+1}\sum_{\nu=1}^{\infty}\frac{g_{\text{N}}(\nu)}{1-g_{\text{N}}(0)}
\left(1-\bar{F}_{\mathcal{P},e}(\nu\xi)-\int_{\nu\xi}^{1}\frac{(\nu\xi)^n}{s^n}\bar{F}_{\mathcal{P},e}(\text{d}s)\right)\frac{(\theta)^n}{n-\delta}{}_{2}F_1\left(n,n-\delta;n-\delta+1;-\theta\right)
\end{split}
\end{equation}
\end{figure*}
\begin{figure*}
\begin{equation}\label{v2}
\begin{split}
V_2(\theta,b)=\delta\sum_{n=0}^{\infty}\binom{b}{n}(-1)^{n+1}\sum_{\nu=1}^{\infty}\frac{g_{\text{N}}(\nu)}{1-g_{\text{N}}(0)}
\left(1-\bar{F}_{\mathcal{P},e}(\nu\xi)-\int_{\nu\xi}^{1}\frac{(\nu\xi)^n}{s^n}\bar{F}_{\mathcal{P},e}(\text{d}s)\right)\frac{\theta^nR^{\alpha n}}{n-\delta}{}_{2}F_1\left(n,n-\delta;n-\delta+1;-\theta R^{\alpha}\right)
\end{split}
\end{equation}
\end{figure*}
\end{Lemma}

\begin{Remark}
When all the users are located within cell-edge region, the users tend to be influenced by each other due to their poor ability to handle the incoming packets. Hence, the derived results provide a lower bound of the network performance. In the scenario where different frequency bands are reserved for CCUs and CEUs. According to the derived results in (\ref{ccu-probability}), the CEUs are distributed as a thinned PPP with density $q(1-R^2)$ \cite{stochastic-geometry-1}. By replacing $q$ with $q(1-R^2)$ in (\ref{ceu-md}), we can obtain the meta distribution for the CEUs in the scenario where different frequency bands are reserved for CCUs and CEUs. The meta distribution of the CCUs can be obtained utilizing a similar recursive approach with that of the CEUs, which is shown as follows
\begin{equation}
\bar{F}_{\mathcal{P},e}=\lim_{t\rightarrow\infty}\bar{F}_{\mathcal{P},e,t}.
\end{equation}
\end{Remark}

\subsection{Mean Local Delay}
The mean local delay is defined as the mean number of transmission attempts until the first success. When the network parameters, i.e., the active probability and the SINR threshold, are below specific thresholds, the according local delay is finite. In other words, the fraction of the BSs suffering from large delays is negligible. However, the local delay could be infinite when the network parameters exceed the threshold, which indicate that a significant number of BSs suffer from large delays. Therefore, we can obtain the mean local delay in the following lemma.
\begin{Lemma}
The mean local delay for a CCU is given by
\begin{equation}
\begin{split}
&M_{-1,c}=\\
&\left(1+\frac{q\delta\theta R^{\alpha}}{1-\delta}{}_2F_1\left(1,1-\delta;2-\delta;-(1-q)\theta R^{\alpha}\right)\right)^{-1}
\end{split}
\end{equation}
\end{Lemma}
\emph{Proof:} See the Appendix D.

Following the similar methodology with the CCU case, the mean local delay for a CEU is given by
\begin{equation}
\begin{split}
&M_{-1,e}=\frac{qR^2}{1-R^2}\sum_{t=0}^{\infty}\sum_{l=0}^{1}\sum_{m=0}^{1}(-1)^{l+m+1}\left((t+m)(1-R^2)\right.\\
&\left.+|m-1|R^2V_1(\theta,-1)+(1-|m-1|)V_2(\theta,-1)+R^{2l}\right)^{-1}\\
&+\frac{1-q}{1-R^2}\left(\frac{1}{1+V_1(\theta,-1)}-\frac{R^2}{1+R^2V_1(\theta,-1)}\right)
\end{split}
\end{equation}
where
\begin{equation}
V_1(\theta,-1)=-\frac{q\theta\delta}{1-\delta}{}_{2}F_1\left(1,1-\delta;2-\delta;-(1-q)\theta\right)
\end{equation}
\begin{equation}
V_2(\theta,-1)=-\frac{q\theta\delta R^{\alpha}}{1-\delta}{}_{2}F_1\left(1,1-\delta;2-\delta;-(1-q)\theta R^{\alpha}\right)
\end{equation}

Next, we provide the threshold of the network parameters which separate the regions of infinite and finite mean local delay. The threshold is also called the critical value. In addition, we say a phase transition occurs when the mean local delay changes from finite to infinite. Note that the critical values for CCU and CEU are different. The critical value of the active probability for a CCU can be obtained by solving the following inequality:
\begin{equation}
\frac{q\theta\delta R^{\alpha}}{1-\delta}{}_{2}F_1\left(1,1-\delta;2-\delta;-(1-q)\theta R^{\alpha}\right)=1.
\end{equation}

Similarly, the critical value of the active probability for a CEU can be obtained by solving the following inequality:
\begin{equation}
\frac{q\theta\delta}{1-\delta}{}_{2}F_1\left(1,1-\delta;2-\delta;-(1-q)\theta\right)=1.
\end{equation}

\section{Simulation Results}
In this section, we fist compare the meta distribution and the beta approximation to demonstrate the accuracy of the beta approximations. Then the effect of the critical system parameters, i.e., the active probability, BS density and the SIR threshold, on the meta distribution for CCU and CEU is investigated. Moreover, the effect of network parameters on the mean local delay for CCU and CEU is also presented. nless otherwise stated, the parameters are set as follows. The BS transmit power is $P=23$dBm, the BS density is $\lambda=10^{-4}/\text{m}^2$, the user density is $\lambda_u=3\times10^{-4}/\text{m}^2$, the path loss exponent $\alpha=3$ and the ratio threshold $R=0.5$.

\begin{figure}
  \centering
  \includegraphics[width=3.5in]{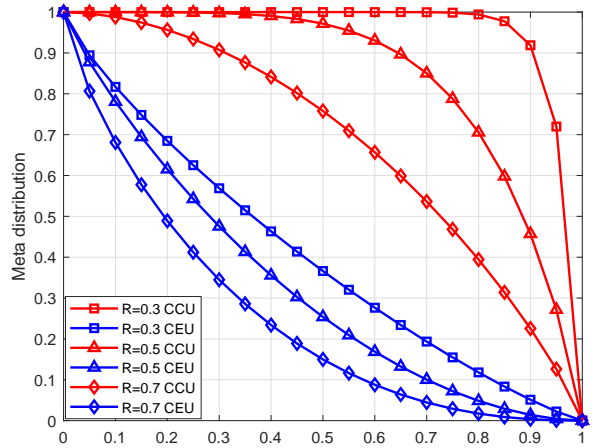}\\
  \caption{Meta distribution under different ratios of between the distance of the user and its serving BS and the dominant interfering BS}\label{meta-distribution-ratio}
\end{figure}

Fig. \ref{meta-distribution-ratio} depicts the meta distribution as a function of the success probability under different ratio thresholds $R$. The meta distribution shows the proportion of CCUs/CEUs achieving different success probability. It can be observed that the meta distribution for CEUs shows a large degradation when the success probability starts to increase, indicating that a relative large proportion of CEUs achieve a low success probability. In contrast, the curves representing the meta distribution for CCUs remain flat in the low success probability regime and decreases rapidly in the large success probability regime. This indicates that a large proportion of CCUs can obtain a larger success probability. Furthermore, the meta distribution of CCUs and CEUs both decrease with the increase of $R$. It can be explained as follows. With a large value of $R$, more users with larger distance from the serving BSs are included into the cell-center region, resulting in the increase of the proportion for the users achieving relatively lower success probabilities. Similarly, the CEUs with relatively larger success probability are extracted from the cell-edge region and the CEUs with lower success probability occupy a larger proportion.

\begin{figure}
  \centering
  \includegraphics[width=3.5in]{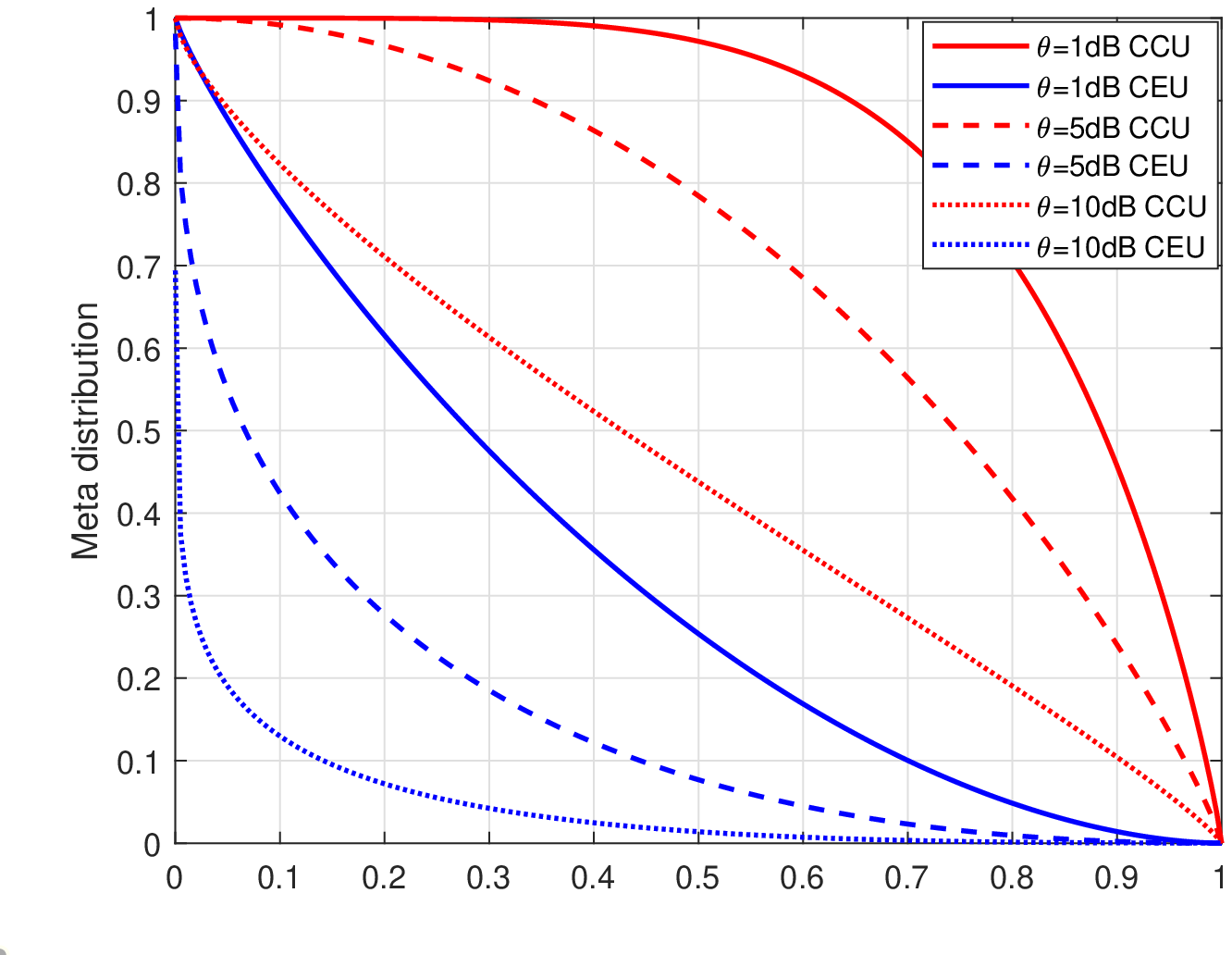}\\
  \caption{Meta distribution under different SIR thresholds}\label{meta-distribution-SIR-threshold}
\end{figure}

Fig. \ref{meta-distribution-SIR-threshold} shows the effect of the SIR threshold $\theta$ on the meta distribution. We can observe that the meta distributions for CCUs and CEUs decrease with $\theta$. The reason is that the the increase of the SIR threshold lead to the decrease of the success probability. Therefore, the proportion of the users with lower success probability increases.

\begin{figure}
  \centering
  \includegraphics[width=3.5in]{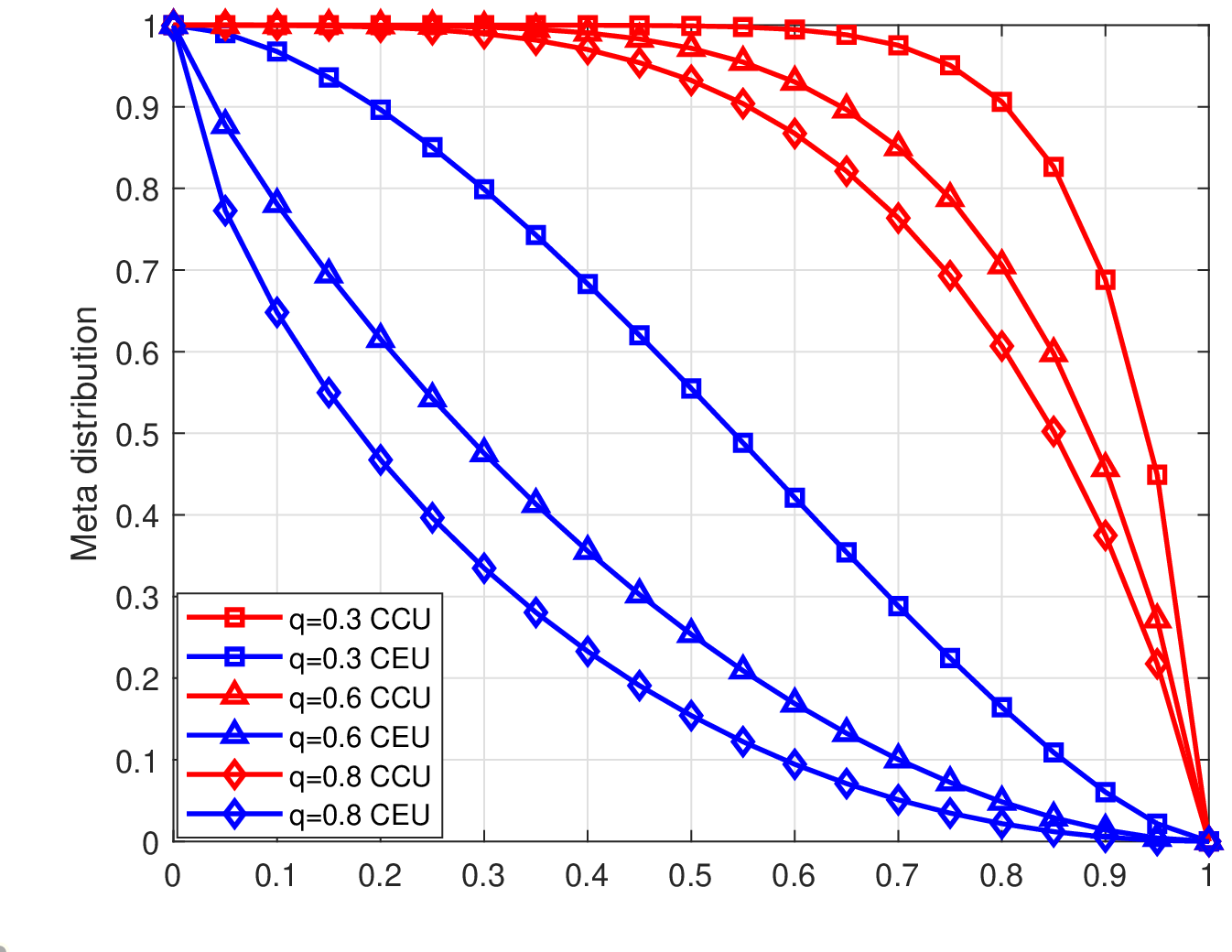}\\
  \caption{Meta distribution under different active probabilities}\label{meta-distribution-active-probability}
\end{figure}

Fig. \ref{meta-distribution-active-probability} shows the effect of the active probability $q$ on the meta distribution for CCUs and CEUs. We can observe that the meta distribution decreases with the active probability, indicating that the proportion of users with lower success probability increases.

\begin{figure}
  \centering
  \includegraphics[width=3.5in]{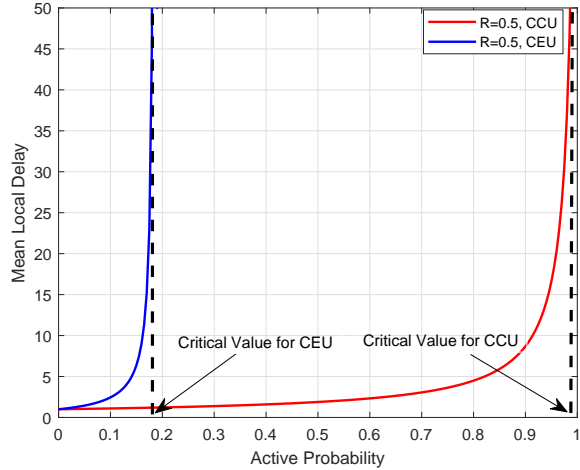}\\
  \caption{Mean local delay under different ratio thresholds}\label{local-delay-active-ratio}
\end{figure}

Fig. \ref{local-delay-active-ratio} illustrates the mean local delay as a function of the active probability. It can be observed that the mean local delays for CCU and CEU increase with the active probability. When the curves representing both local delays reach the critical value of the active probability, the mean local delay approach infinite. Note that the critical value for CCU is far larger than that for CEU. This coincides with our intuition. The distance between CEU and its tagged BS is larger, resulting in a lower service rate. Therefore, the mean local delay tends to be larger.

\begin{figure}
  \centering
  \includegraphics[width=3.5in]{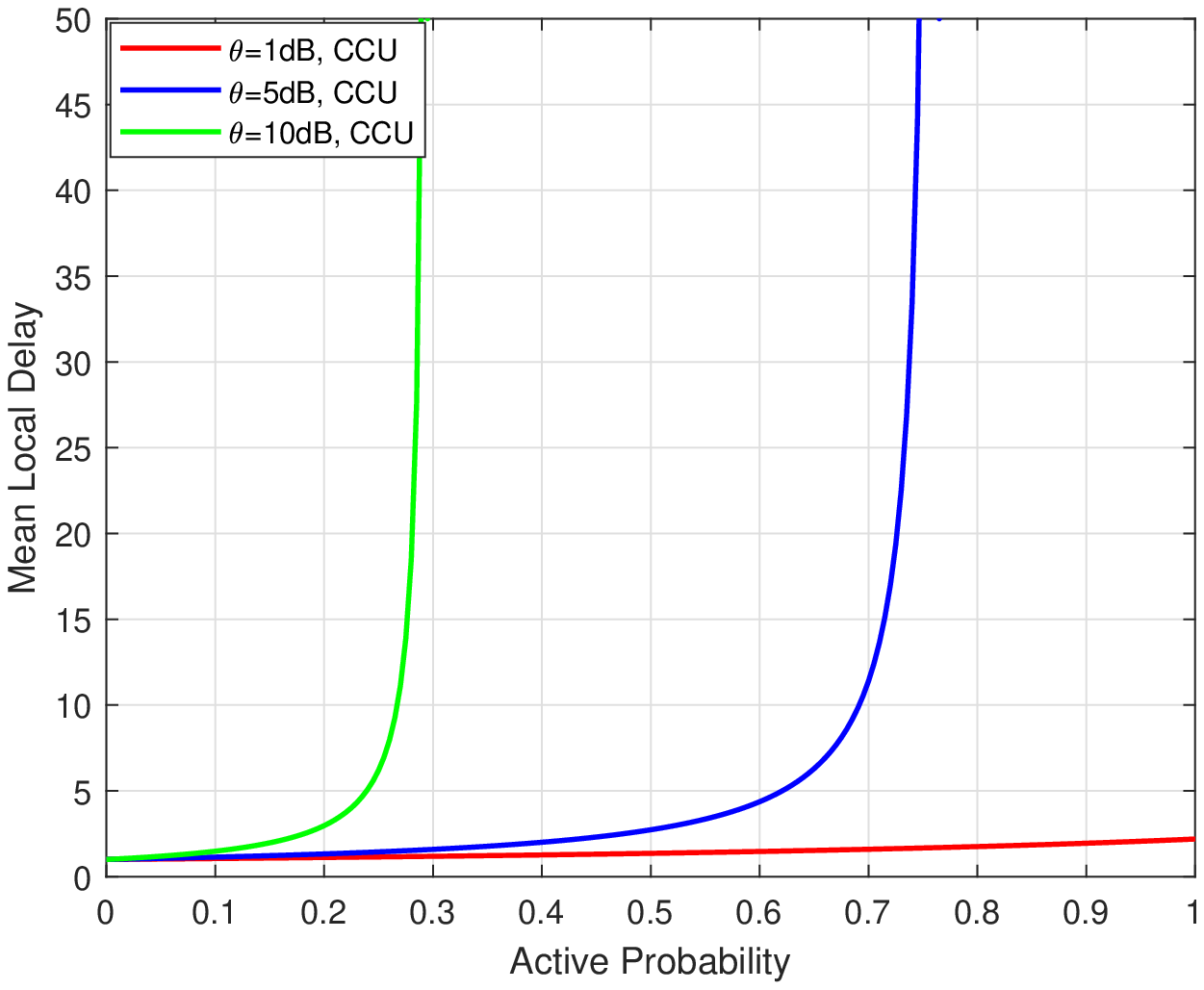}\\
  \caption{Mean local delay under different SIR thresholds}\label{local-delay-active-SIR-threshold}
\end{figure}

\begin{figure}
  \centering
  \includegraphics[width=3.5in]{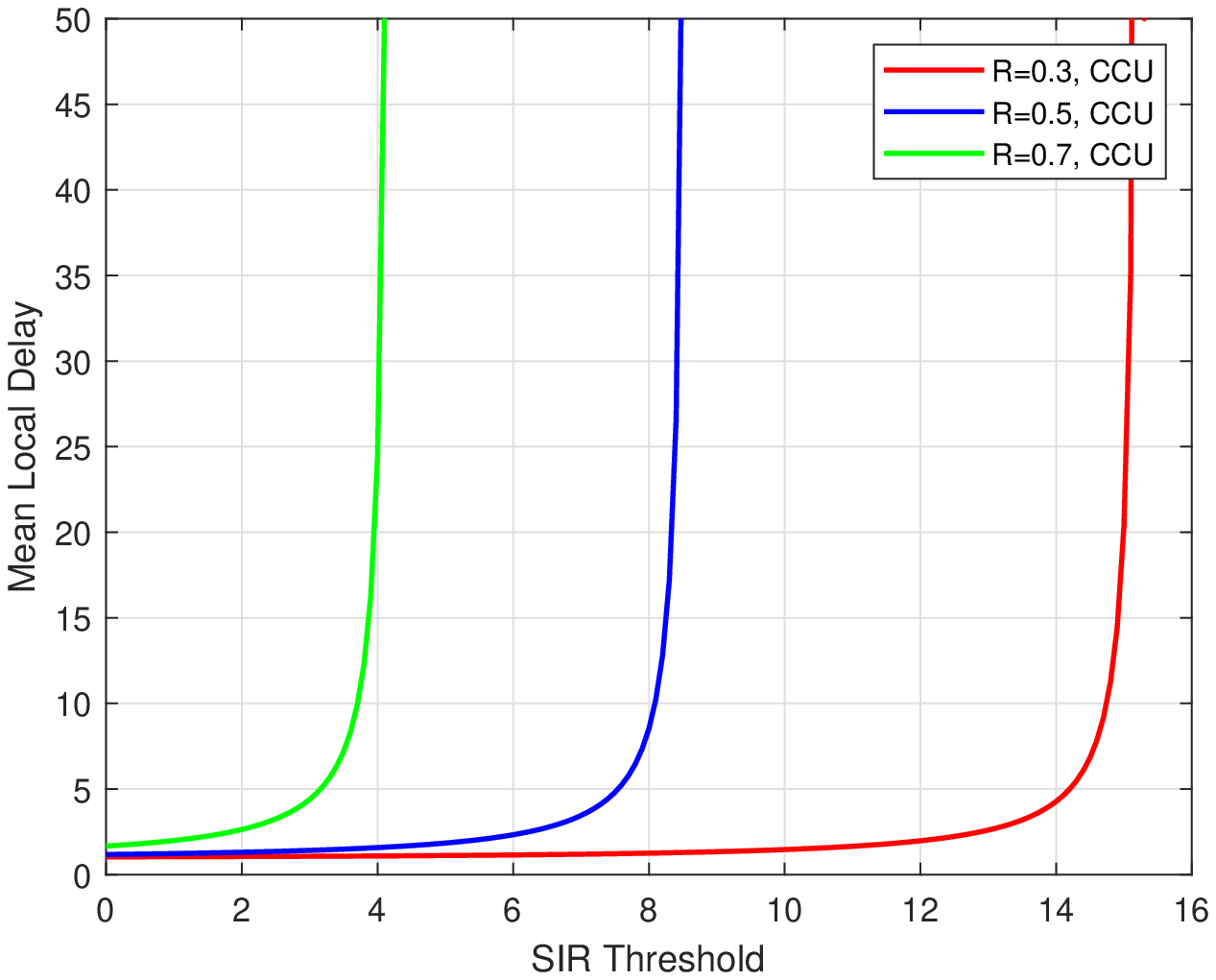}\\
  \caption{Mean local delay under different SIR thresholds}\label{local-delay-SIR-threshold-ratio}
\end{figure}

Fig. \ref{local-delay-active-SIR-threshold} and Fig. \ref{local-delay-SIR-threshold-ratio} illustrates the mean local delay as a function of the active probability under various SIR thresholds. It can be observed that the increasing SIR threshold will enlarge the mean local delay. When the SIR threshold $\theta=1$dB, the mean local delay maintain finite for all the active probabilities. However, when $\theta$ increases to 5dB, the mean local delay goes to infinity when the active probability is approximately 0.7. In addition, the mean local delay reaches infinity at a lower value of active probability when $\theta$ becomes larger.

From Fig. \ref{local-delay-SIR-threshold-ratio}, we can see that the the mean local delay reaches infinity  when the increasing ratio threshold $R$ also increases the critical value of the SIR threshold.

\begin{figure}
  \centering
  \includegraphics[width=3.5in]{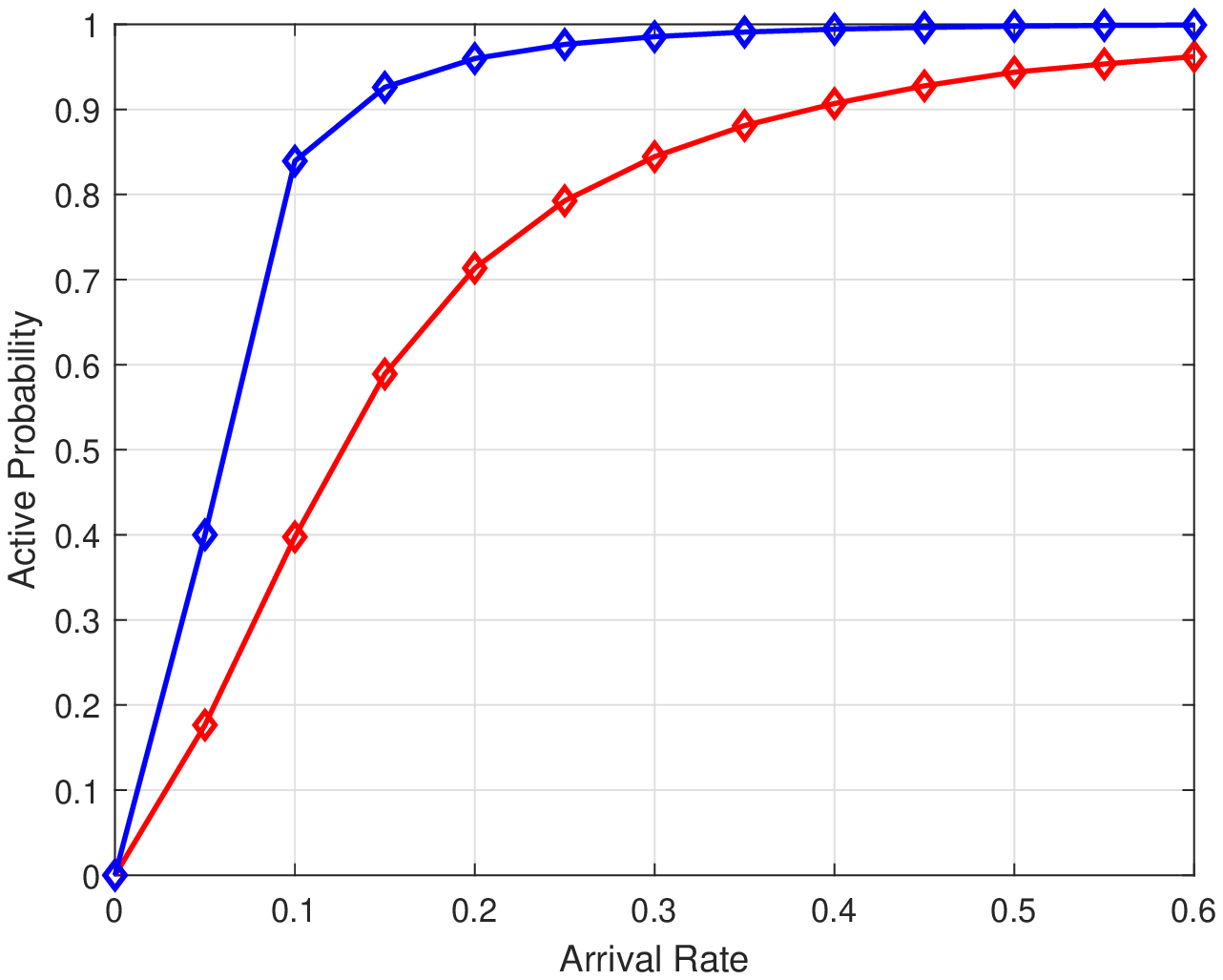}\\
  \caption{active probability vs arrival rate}\label{active-probability-arrival-rate}
\end{figure}

\begin{figure}
  \centering
  \includegraphics[width=3.5in]{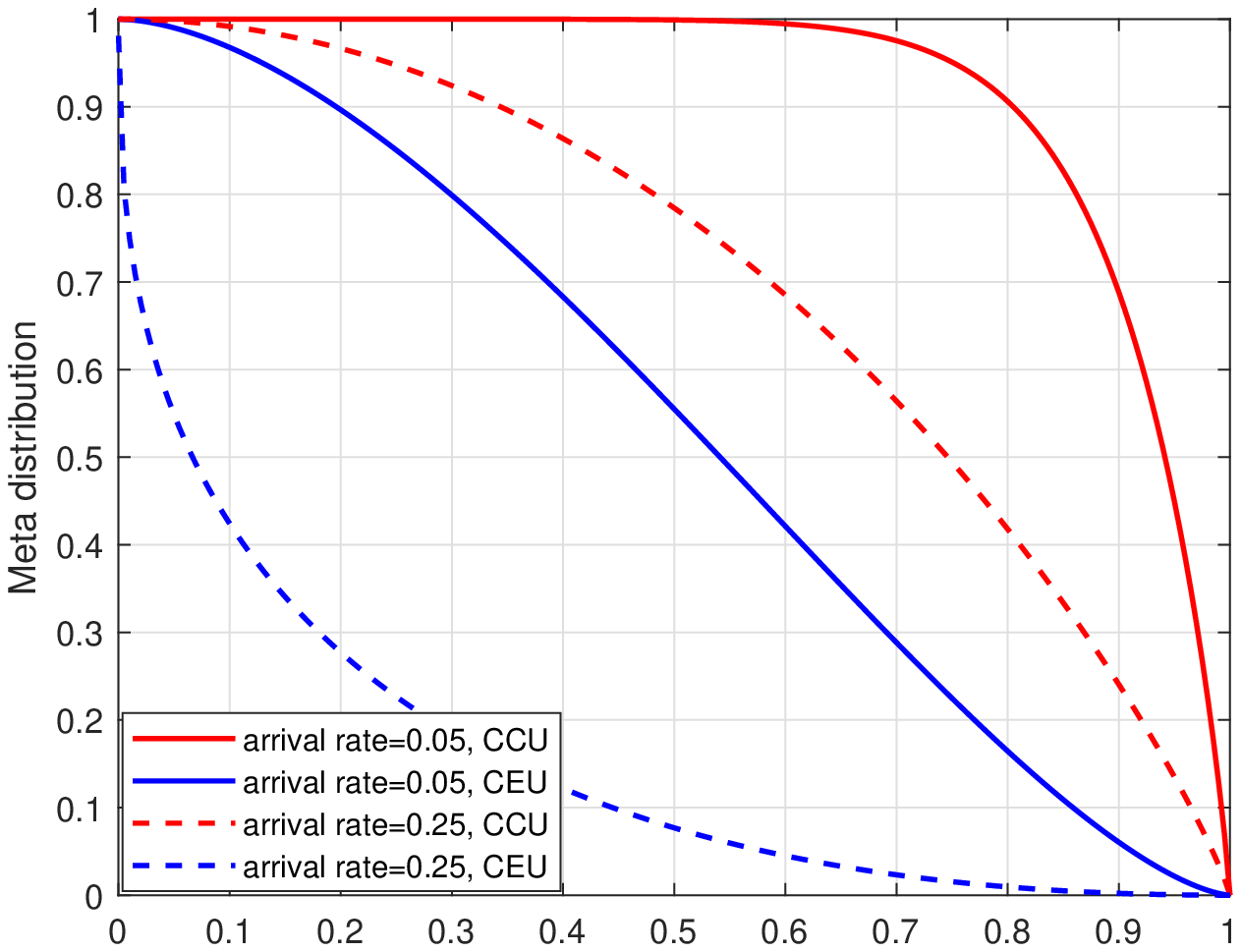}\\
  \caption{Mean local delay under different SIR thresholds}\label{meta-distribution-SIR-arrival-rate}
\end{figure}

Fig. \ref{active-probability-arrival-rate} illustrates the relationship between the active probability and the arrival rate. It can be observed the active probabilities of the CCUs and CEUs both increases with the arrival rates. When the arrival rate for CCUs and CEUs is relatively low, the active probability increases with the arrival rate rapidly. Specifically, the active probability for CEUs exceed 0.8 when the arrival rate is 0.1. This shows that the CEUs are   Afterwards, the active probability grows slower. Fig. \ref{meta-distribution-SIR-arrival-rate} shows the effect of the arrival rates on the meta distribution for the CCUs and CEUs. It is notified that the number of CCUs and CEUs achieving the lower success probability increases significantly when the traffic load varies from light ($\xi$=0.01) to heavy ($\xi$=0.25).

\section{Conclusion}
we develop a mathematical analytical model by utilizing queuing theory and stochastic geometry where the randomness of the traffic and the geographical locations of the interferers can be captured. We then derive the closed-form expressions of the meta distribution for the cell-center users (CCUs) and the cell-edge users (CEUs), respectively. Fixed-point equations are formulated to obtain the exact value of the meta distribution by taking the random arrival traffic into consideration and the impact of the random arrival traffic on the queue status is revealed. In addition, the mean local delay is derived and the corresponding region where the mean local delay maintain finite is obtained.

\section{Appendix}
\subsection{Proof of Theorem 1}

The success probability conditioned on $\Phi$ when a typical CCU is served by a BS with link distance $R_c$ is
\begin{equation}
\begin{split}
&\mathbbm{P}(\text{SIR}>\theta|\Phi)\\
&=\mathbbm{P}\left(\left.\frac{P\left|h_{r_c}\right|^2r_c^{-\alpha}}{\sum_{x\in\Phi\backslash B\left(0,\frac{r_c}{R}\right)}\mathbbm{1}\left(x\in\Phi(t)\right)Ph_xx^{-\alpha}}>\theta\right|\Phi\right)\\
&\overset{(a)}{=}\mathbbm{E}\left[\exp\left(-\frac{\theta r_c^{\alpha}}{P}\sum_{x\in\Phi\backslash B\left(0,\frac{r_c}{R}\right)}\mathbbm{1}\left(x\in\Phi(t)\right)P\left|h_{r_c}\right|^2r_c^{-\alpha}\right)\right]\\
&=\prod_{x\in\Phi\backslash B\left(0,\frac{r_c}{R}\right)}\left(\frac{q}{1+\theta r_c^{\alpha}x^{-\alpha}}+1-q\right),
\end{split}
\end{equation}
where (a) follows $\left|h_{r_c}\right|^2\sim\exp(1)$.
Accordingly, the $b$-th moment of the conditional success probability for a CCU is
\begin{equation}
\begin{split}
&M_{b,c}=\mathbbm{E}_{\Phi}\left[\mathbbm{P}\left(\text{SIR}>\theta|\Phi\right)^{b}\right]\\
&=\mathbbm{E}_{\Phi}\left[\prod_{x\in\Phi\backslash B\left(0,\frac{r_c}{R}\right)}\left(\frac{q}{1+\theta r_c^{\alpha}x^{-\alpha}}+1-q\right)^b\right]\\
&\overset{(a)}{=}\mathbbm{E}_{\Phi}\left[\exp\left(-2\pi\lambda\int_{\frac{r_c}{R}}^{\infty}\right.\right.\\
&\left.\left.\left(1-\left(\frac{q}{1+\theta r_c^{\alpha}x^{-\alpha}}+1-q\right)^b\right)x\text{d}x\right)\right]\\
&=\mathbbm{E}_{\Phi}\left[\exp\left(-\pi\delta\lambda\sum_{n=1}^{\infty}\binom{b}{n}(-1)^{n+1}
\frac{(q\theta)^nr_c^2R^{\alpha n}}{R^2(n-\delta)}\right.\right.\\
&\left.\left.{}_{2}F_{1}\left(n,n-\delta;n-\delta+1;-\theta R^{\alpha}\right)\right)\right]\\
&\overset{(b)}{=}\frac{2\pi\lambda}{R^2}\int_{0}^{\infty}r_c\exp\left(-\pi\lambda\frac{r_c^2}{R^2}-\pi\delta\lambda\sum_{n=1}^{\infty}\binom{b}{n}(-1)^{n+1}\right.\\
&\left.\frac{(q\theta)^nr_c^2R^{\alpha n}}{R^2(n-\delta)}{}_{2}F_{1}\left(n,n-\delta;n-\delta+1;-\theta R^{\alpha}\right)\right)\text{d}r_c\\
&=\left(1+\delta\sum_{n=1}^{\infty}\binom{b}{n}(-1)^{n+1}\frac{(q\theta)^nR^{\alpha n}}{n-\delta}\right.\\
&\left.{}_{2}F_1\left(n,n-\delta;n-\delta+1;-\theta R^{\alpha}\right)\right)^{-1},
\end{split}
\end{equation}
where (a) is obtained by using the probability generating functional (PGFL) of a PPP \cite{slivnyak}, and (b) follows from (\ref{ccu-pdf}).

\subsection{Proof of Lemma 2}

Let $I_{e1,1}$ and $I_{e1,2}$ be the interferences from the BSs in the set $\mathcal{S}_1(r_e)$ and $\mathcal{S}_2(r_e)$, i.e., $I_{e1,1}=\sum_{x\in\mathcal{S}_1(r_e)}\mathbbm{1}\left(x\in\Phi(t)\right)Ph_xx^{-\alpha}$ and $I_{e1,2}=\sum_{x\in\mathcal{S}_2(r_e)}\mathbbm{1}\left(x\in\Phi(t)\right)Ph_xx^{-\alpha}$, respectively. Thus, the success probability conditioned on $\Phi$ when a typical CEU is served by a BS with link distance $r_e$ is
\begin{equation}
\begin{split}
\mathbbm{P}(\text{SIR}>\theta|\Phi)=&\mathbbm{P}\left(\left.\frac{P\left|h_{r_e}\right|^2r_e^{-\alpha}}{I_{e1,1}+I_{e1,2}}>\theta\right|\Phi\right)\\
=&\mathbbm{E}\left[\exp\left(-\frac{\theta r_e^{\alpha}}{P}\left(I_{e1,1}+I_{e1,2}\right)\right)\right]\\
\overset{(a)}{=}&\prod_{x\in\mathcal{S}_1(r_e)}\left(\frac{q}{1+\theta r_e^{\alpha}x^{-\alpha}}+1-q\right)\\
&\prod_{x\in\mathcal{S}_2(r_e)}\left(\frac{q}{1+\theta r_e^{\alpha}x^{-\alpha}}+1-q\right),
\end{split}
\end{equation}
where (a) is obtained by utilizing the moment generating function (MGF) of $\left|h_{r_c}\right|^2$.

Next, we derive the $b$-th moment of conditional success probability for $\mathcal{S}_1(r_e)$ and $\mathcal{S}_2(r_e)$. Let $M_{b,e1,1}$ and $M_{b,e1,2}$ be the $b$-th moment of the conditional success probability for a CEU corresponding to the set $\mathcal{S}_1(r_e)$ and $\mathcal{S}_2(r_e)$, respectively. Thus, $M_{b,e1}$ is the addition of $M_{b,e1,1}$ and $M_{b,e1,2}$. Since the derivation of $M_{b,e1,2}$ is similar with that for CCU, we first derive $M_{b,e1,2}$ conditioned on $\Phi$ as follows
\begin{equation}\label{M-e1-2}
\begin{split}
&M_{b,e1,2}(\theta|\Phi)=\mathbbm{P}\left(\text{SIR}>\theta|\Phi\right)^{b}\\
&=\prod_{x\in\Phi\backslash B\left(0,\frac{r_e}{R}\right)}\left(\frac{q}{1+\theta r_e^{\alpha}x^{-\alpha}}+1-q\right)^b\\
&=\exp\left(-\pi\delta\lambda\sum_{n=1}^{\infty}\binom{b}{n}(-1)^{n+1}
\frac{(q\theta)^nr_e^2R^{\alpha n}}{R^2(n-\delta)}\right.\\
&\left.{}_{2}F_{1}\left(n,n-\delta;n-\delta+1;-\theta R^{\alpha}\right)\right),
\end{split}
\end{equation}

Now, we derive $M_{b,e1,1}$. Note that in the PPP with density $\lambda$, the number of nodes in the area $A$ is a Poisson distributed random variable with mean $\lambda A$ and each node is uniformly distributed in the area $A$. Thus, the number of nodes in the set $\mathcal{S}_1$ follows Poisson distribution with mean $\pi\lambda r_e^2\left(R^{-2}-1\right)$ and each node is distributed as
\begin{equation}
f(x)=
\begin{cases}
\frac{1}{\pi r_e^2(R^{-2}-1)},\,r_e\leq\left\|x\right\|\leq\frac{r_e}{R},\\
0,\ \ \ \ \ \ \ \ \ \ \ \ \,\text{otherwise}.
\end{cases}
\end{equation}

We denote the number of nodes in the set $\mathcal{S}_1(r_e)$ by $K$. Conditioned on the existence of at least one interfering BS in $\mathcal{S}_1(r_e)$, the MGF of $K$ can be derive as follows
\begin{equation}\label{K-MGF}
\begin{split}
\mathbbm{E}\left[z^k|k\geq1\right]&\overset{(a)}{=}\frac{\mathbbm{E}[z^k]-\exp\left(-\pi\lambda r_e^2(R^{-2}-1)\right)}{1-\exp\left(-\pi\lambda r_e^2(R^{-2}-1)\right)}\\
&\overset{(b)}{=}\frac{\exp\left(\mathcal{C}(r_e)(1-z)\right)-\exp\left(\mathcal{C}(r_e)\right)}{1-\exp\left(\mathcal{C}(r_e)\right)},
\end{split}
\end{equation}
where (a) follows from the void probability of a PPP, (b) is from the MGF of a Poisson distributed random variable and $\mathcal{C}(r_e)=\pi\lambda r_e^2(R^{-2}-1)$.

Based on the above analysis, we derive $M_{b,e1,1}$ conditioned on $\Phi$ as follows
\begin{equation}\label{M-e1-1}
\begin{split}
&M_{b,e1,1}(\theta|\Phi)=\mathbbm{P}\left(\text{SIR}>\theta|\Phi\right)^{b}\\
&=\prod_{x\in\mathcal{S}_1(r_e)}\left(\frac{q}{1+\theta r_c^{\alpha}x^{-\alpha}}+1-q\right)^b\\
&\overset{(a)}{=}\mathbbm{E}_k\left[\left.\left(2\pi\int_{r_e}^{\frac{r_e}{R}}\left(\frac{q}{1+\theta r_c^{\alpha}x^{-\alpha}}+1-q\right)^bxf(x)\text{d}x\right)^k\right|k\geq1\right]\\
&\overset{(b)}{=}\frac{1}{1-\exp\left(-\mathcal{C}(r_e)\right)}\left(\exp\left(-\mathcal{C}(r_e)\int_{r_e}^{\frac{r_e}{R}}\right.\right.\\
&\left.\left.\left(1-\left(\frac{q}{1+\theta r_e^{\alpha}x^{-\alpha}}+1-q\right)^b\right)f(x)\text{d}x
\right)-\exp\left(-\mathcal{C}(r_e)\right)\right)\\
&=\frac{1}{1-\exp\left(-\mathcal{C}(r_e)\right)}\left(\exp\left(-2\pi\lambda\int_{r_e}^{\frac{r_e}{R}}\right.\right.\\
&\left.\left.\left(1-\left(\frac{q}{1+\theta r_c^{\alpha}x^{-\alpha}}+1-q\right)^b\right)x\text{d}x
\right)-\exp\left(-\mathcal{C}(r_e)\right)\right)\\
&\overset{(c)}{=}\frac{\exp\left(\pi\lambda r_e^2R^{-2}V_2(\theta,b)-\pi\lambda r_e^2V_1(\theta,b)\right)-\exp\left(-\mathcal{C}(r_e)\right)}{1-\exp(-\mathcal{C}(r_e))}
\end{split}
\end{equation}
where (a) follows from the i.i.d. property of each node in a PPP, (b) is from (\ref{K-MGF}). In addition, $V_1(\theta,b)$ and $V_2(\theta,b)$ is given by (\ref{V-1}) and (\ref{V-2}), respectively.

Since we have obtain the $b$-th moment of the conditional success probability conditioned on $\Phi$ corresponding to $\mathcal{S}_1(r_e)$ and $\mathcal{S}_2(r_e)$, the $b$-th moment of the conditional success probability for a CEU when the dominant interfering BS is active can be derived as follows
\begin{equation}\label{M-e1}
\begin{split}
M_{b,e1}&=\mathbbm{E}_{\Phi}\left[M_{b,e1,1}(\theta|\Phi)M_{b,e1,2}(\theta|\Phi)\right]\\
&=\int_{0}^{\infty}M_{b,e1,1}(\theta|\Phi)M_{b,e1,2}(\theta|\Phi)f_{R_e}(r_e)\text{d}r_e
\end{split}
\end{equation}

Substituting (\ref{ceu-pdf}) (\ref{M-e1-2}) and (\ref{M-e1-1}) into (\ref{M-e1}) yields (\ref{M-e1-derivation}), which is shown at the bottom of the next page.
\begin{figure*}
\begin{equation}\label{M-e1-derivation}
\begin{split}
M_{b,e1}=&\frac{2\pi\lambda}{1-R^2}\int_{0}^{\infty}
\frac{\exp\left(\pi\lambda r_e^2R^{-2}V_2(\theta,b)-\pi\lambda r_e^2V_1(\theta,b)\right)-\exp\left(-\mathcal{C}(r_e)\right)}{1-\exp(-\mathcal{C}(r_e))}\exp\left(-\pi\lambda r_e^2R^{-2}V_2(\theta,b)\right)\\
&\left(\exp\left(-\pi\lambda r_e^2\right)-\exp\left(-\pi\lambda\frac{r_e^2}{R^2}\right)\right)r_e\text{d}r_e\\
&\overset{(a)}{=}\frac{1}{n+(1+V_1(\theta,b)-n)R^2}-\frac{1}{1+n+(V_1(\theta,b)-n)R^2}+\frac{1}{n+2+V_2(\theta,b)-(1+n)R^2}-\frac{1}{1+n+V_2(\theta,b)-nR^2}
\end{split}
\end{equation}
\end{figure*}
where (a) is obtained by utilizing Maclaurin series $\frac{1}{1-x}=\sum_{0}^{\infty}x^n$. Rearranging the solution of integral
yields (\ref{M-e1}). Then the proof is completed.

\subsection{Proof of Lemma 3}

Conditioned on that the dominant interfering BS is not active, the $b$-th moment of the conditional success probability can be derived as follows
\begin{equation}\label{M-e2}
\begin{split}
&M_{b,e2}=\mathbbm{E}_{\Phi}\left[\mathbbm{P}\left(\text{SIR}>\theta|\Phi\right)^{b}\right]\\
&=\mathbbm{E}_{\Phi}\left[\prod_{x\in\Phi\backslash B\left(0,\frac{r_e}{R}\right)}\left(\frac{q}{1+\theta r_c^{\alpha}x^{-\alpha}}+1-q\right)^b\right]\\
&=\mathbbm{E}_{\Phi}\left[\exp\left(-\pi\delta\lambda\sum_{n=1}^{\infty}\binom{b}{n}(-1)^{n+1}
\frac{(q\theta)^nr_e^2}{(n-\delta)}\right.\right.\\
&\left.\left.{}_{2}F_{1}\left(n,n-\delta;n-\delta+1;-\theta\right)\right)\right]\\
&=\int_{0}^{\infty}\exp\left(-\pi\delta\lambda\sum_{n=1}^{\infty}\binom{b}{n}(-1)^{n+1}
\frac{(q\theta)^nr_e^2}{(n-\delta)}\right.\\
&\left.{}_{2}F_{1}\left(n,n-\delta;n-\delta+1;-\theta\right)\right)f_{R_e}(r_e)\text{d}r_e\\
&=\frac{1}{1-R^2}\left(\frac{1}{1+V_1(\theta,b)}-\frac{R^2}{1+R^2V_1(\theta,b)}\right),
\end{split}
\end{equation}
where $f_{R_e}(r_e)$ and $V_1(\theta)$ are given by (\ref{ceu-pdf}) and (\ref{V-1}), respectively.

\subsection{Proof of Lemma 4}

The mean local delay can be derived as follows
\begin{equation}
\begin{split}
&M_{-1,e}=\mathbbm{E}_{\Phi}\left[\mathbbm{P}\left(\text{SIR}>\theta|\Phi\right)^{-1}\right]\\
&=\mathbbm{E}_{\Phi}\left[\prod_{x\in\Phi\backslash B\left(0,\frac{r_c}{R}\right)}\left(\frac{q}{1+\theta r_c^{\alpha}x^{-\alpha}}+1-q\right)^{-1}\right]\\
&=\mathbbm{E}_{\Phi}\left[\exp\left(2\pi\lambda\int_{\frac{r_c}{R}}^{\infty}\frac{q\theta r_c^{\alpha}x^{1-\alpha}}{1+(1-q)\theta r_c^{\alpha}x^{-\alpha}}\text{d}x\right)\right]\\
&=\mathbbm{E}_{\Phi}\left[\exp\left(-\pi\delta\lambda\sum_{n=1}^{\infty}\binom{b}{n}(-1)^{n+1}
\frac{q\theta r_c^2}{(1-\delta)}\right.\right.\\
&\left.\left.{}_{2}F_{1}\left(1,1-\delta;2-\delta;-(1-q)\theta\right)\right)\right]\\
&=\int_{0}^{\infty}\exp\left(-\pi\delta\lambda\sum_{n=1}^{\infty}\binom{b}{n}(-1)^{n+1}
\frac{ q\theta r_c^2}{(1-\delta)}\right.\\
&\left.{}_{2}F_{1}\left(1,1-\delta;2-\delta;-(1-q)\theta\right)\right)f_{R_c}(r_c)\text{d}r_c\\
&=\left(1+\frac{q\delta\theta R^{\alpha}}{1-\delta}{}_2F_1\left(1,1-\delta;2-\delta;-(1-q)\theta R^{\alpha}\right)\right)^{-1},
\end{split}
\end{equation}

\end{spacing}
\end{document}